**Avalanche-like metamagnetic transition in (LaNd)CaMnO manganites**

L. Ghivelder[1,*], G.G. Eslava[1], R.S. Freitas[2], G. Leyva[3,4] and F. Parisi[3,4]

[1] Instituto de Física, Universidade Federal do Rio de Janeiro, Rio de Janeiro, RJ 21941-972, Brazil

[2] Instituto de Física, Universidade de São Paulo, São Paulo, SP 05314-970, Brazil

[3] Departamento de Física, Centro Atómico Constituyentes, Comision Nacional de Energía Atomica, Prov. de Buenos Aires, Argentina

[4] Escuela de Ciencia y Tecnología, Universidad Nacional de General San Martín, Buenos Aires, Argentina

*abstract*

*We investigate the behavior of ultrasharp metamagnetic transitions in $La_{5/8-y}Nd_yCa_{3/8}MnO_3$ manganites. These compounds change from a low temperature ferromagnetic metallic state at low Nd doping to a charge-ordered antiferromagnetic insulator for high Nd content. At an intermediate doping a phase-separated state is established. At low temperatures (~ 2 K), we observe an avalanche-like field-induced metamagnetic transition, when the entire compound changes abruptly from one phase to the other. We investigate the signatures of this* ultrasharp *transition using magnetization and specific heat measurements. We observe a first order transition in the specific heat associated with discontinuous jumps in the magnetization. A strong increase of the sample temperature is simultaneously observed. The results are interpret in terms of latent heat release from the field induced enhancement of the ferromagnetic fraction, triggering the avalanche process.*

# 1 – Introduction

The exotic physical properties of mixed valence manganese oxides, known as manganites, have been fascinating and puzzling the scientific community for the last two decades [1]. The most striking phenomena are the colossal magnetoresistance, charge and orbital ordered states, and the coexistence at different length scales of ferromagnetic (FM) metallic and charge ordered antiferromagnetic (CO-AFM) insulating domains. The latter, known as phase separation, is currently viewed as an intrinsic feature of several strongly correlated electron systems [2], such as superconductors [3], multi-ferroics [4], and magnetocaloric materials [5]. Intrinsic disorder and the coexistence of energetically near degenerate phases are key factors to understand this effect [2,6].

Metamagnetic transitions, observed in measurements of isothermal magnetization versus magnetic field, M vs. H, are common in manganites [7,8]. The origin of these effects are caused by the field induced transformation of the metastable CO-AFM state at low fields to a homogeneous ferromagnet at high fields. In manganites both phases coexist in low magnetic fields, and the broad metamagnetic transitions observed, are the result of an increasing field dependent fraction of the ferromagnetic phase within the phase-separated material. But in addition, measurements at very low temperatures (< 4 K) have revealed a surprising and beautiful effect: the observation of a discontinuous metamagnetic transition at a critical magnetic field [9,10,11]. This phenomenon, referred to as ultrasharp magnetization jumps, avalanche-like or abrupt metamagnetic transition, has puzzled the scientific community since it was first reported [12]. Similar effects were later found on a variety of different systems [13,14], including spin ice [15] and iron-rich intermetallic compounds [16], and it is now established that these discontinuous transitions are not a peculiarity restricted to manganese oxides. More generally, the study of magnetic avalanches in manganites helps the understanding of phase transitions of mixed first order and continuous character, and from the study of this phenomena other fields can benefit, such as fracture theory in disordered media [17].

Several aspects of these discontinuous transition were previously studied. It has been established that the critical field of the avalanche-like transition depends strongly on the magnetic field sweep rate [18,19,20,21]. This is a consequence of relaxation effects, which play a major role in the avalanche process, and yield the observation of a spontaneous transition [10,11,13] at a fixed magnetic field. The main question was, and to some extent still is, what causes these discontinuous magnetization jumps? Initially, an interpretation was proposed within the framework of a martensitic transformation [19,22], associated with strain between the phase separated regions. However, various experimental findings rule out the martensitic scenario, such as the sharpness of the transition in inhomogeneous polycrystalline samples, and the cooling rate dependence of the critical field [18]. The explanation more widely accepted is that the avalanches are a consequence of the frozen disorder within the phase separated state, and the sudden release of an excess entropy is in large part responsible for triggering the magnetization jump [18]. It is also known that one necessary ingredient for the observation of avalanche-like magnetization jumps is the presence of blocked metastable states at low temperatures [8,11,18]. This can be experimentally observed through the zero field cooled magnetization, which must start well below the field cooled data due to frozen quenched disorder. The system is trapped in a complex free energy landscape, similar to that which describes spin glasses [23]. With increasing temperature, the magnetic moments become unblocked, and the magnetization rises when the thermal energy is sufficient to overcome the energy barriers.

This paper investigates the metamagnetic transition in phase separated manganite compounds of the family of $La_{5/8-y}Nd_yCa_{3/8}MnO_3$. It is a solid solution between $La_{5/8}Ca_{3/8}MnO_3$ and $Nd_{5/8}Ca_{3/8}MnO_3$. The La based material is an optimized colossal magnetoresistive compound, a nearly homogeneous metallic ferromagnet at low temperatures. On the other hand, the Nd based system exhibits, at low temperatures, a robust charge-ordered insulating antiferromagnetic state. A mixture of both compounds yield the formation of a phase-separated state [24,25]. The $Mn^{4+}/Mn^{3+}$ ratio is kept constant, the only variable being the average A-site ionic radius and the associated disorder. In comparison with the most studied manganite compounds $La_{5/8-y}Pr_yCa_{3/8}MnO_3$, $Nd^{+3}$ ions are slightly smaller than $Pr^{+3}$. Therefore, when doping the La manganite with Nd, the effect of destabilizing the FM state in favor of the AFM-CO ground state are enhanced due a smaller tolerance factor, or in other words, greater Mn–O–Mn distortions when compared to the Pr doped compound. Our study focuses on the metamagnetic transition at low temperatures, and the identification of key factors which determine the occurrence of the avalanche process.

## 2 – Experimental

Bulk polycrystalline samples of $La_{5/8-y}Nd_yCa_{3/8}MnO_3$ were synthesized following the liquid mix technique, starting from the metal carbonates. High resolution x-ray powder diffraction (XPD) measurements were conducted at the Brazilian Synchrotron Light Laboratory (LNLS), with a wavelength of 1.3777 Å, using a linear detector. Magnetization and specific heat were measured using a Quantum Design PPMS system. With the PPMS platform, and under high vacuum, we have also measured directly the sample's temperature rise while the field is increased. Additional magnetization measurements were made with a home assembled VSM magnetometer, mounted in a pumped $He^4$ cryostat, a system designed to allow the sample to be immersed in either liquid helium or gas.

## 3 – Results and Discussion

The inset of Fig. 1 shows room temperature x-ray diffraction results in $La_{0.125}Nd_{0.5}Ca_{3/8}MnO_3$ (y = 0.5). Rietveld analysis was performed using an orthorhombic *Pnma* space group. All diffraction peaks were accounted for, with no impurity phases within the precision of the measurements. The values of the reliability factors of the fitting are $R_{Bragg}$ = 11.2%, $R_F$ = 9.7% and $\chi^2$ = 2.37. The main panel of Fig. 1 shows the room temperature cell parameters, as determined for the entire range of Nd doping in the compounds. There is a smooth and rather small variation of the dimensions of the unit cell, indicating that Nd doping induces no substantial changes in the crystallographic structure. As expected, replacing $La^{3+}$ (*r* = 1.216 Å) with $Nd^{3+}$ (*r* = 1.163 Å), reduces the mean ionic radius of the A-site, as well as the volume of unit cell. This decrease is followed by an increase tilting of $MnO_6$ octahedron, which occurs when additional room is created inside the unit cell [26]. This is confirmed by our Rietveld analysis, which yield an Mn-O-Mn bond angle changing from 164° in the sample with y = 0.1, to 155° for the y = 0.5 compound.

Figure 2 shows the zero field cooled (ZFC) and field cooled (FC) magnetization as a function of temperature, M vs. T, of $La_{5/8-y}Nd_yCa_{3/8}MnO_3$ for y = 0.3, 0.4, and 0.5, measured with a low field, H = 0.2 T. For the highest Nd doped sample, y = 0.5, the charge ordered and antiferromagnetic transition appear at 220 and 180 K, respectively. The CO transition is also clearly visible in the data for the other samples, y = 0.3 and 0.4, and in addition a ferromagnetic phase develops at intermediate temperatures. The low temperature magnetization decreases as a function of the Nd content, indicating that Nd doping inhibits the formation of ferromagnetism. The small hysteresis between the ZFC and FC curves, observed in the samples with Nd 0.4 and 0.5 content, signals the onset of low temperature blocked states, identified as one of the main ingredients for the occurrence of abrupt metamagnetic transitions at low temperatures [11]. We further investigate this issue by measuring the temperature dependence of the magnetization of the y = 0.5 sample, with applied fields from 1 to 4 T, as shown in Fig 3. With a moderate field of 1 T (Fig. 3a) the CO and AFM transitions are evident, and little FM phase is formed down to low temperatures. With increasing field, FM regions start to grow at temperatures below 100 K. This favors the appearance of the phase separated state, and a large difference between ZFC and FC curves develops, typical of a system with metastable states. The overall results of Fig. 3 are not observed in the majority of phase separated manganites. More commonly, block states and hysteresis are present at low fields, and disappear as the field increases above a few Tesla [27].

The metamagnetic transition is investigated in Fig. 4, which shows the magnetization as a function of applied field, M vs. H, measured at 2.5 K in compounds with varying Nd doping. In the y = 0.5 sample, the magnetization remains low up to H = 4.7 T, where a magnetization jump is observed, and the sample abruptly changes to homogenous FM state. This FM state is preserved when the field is lowered, and can only be destroyed once the sample is warmed to higher temperatures. The y = 0.4 sample shows a much higher saturation of the magnetization at fields below 2.6 T, where a magnetization jump also occurs. The magnetization values below and above the metamagnetic transition indicate that the FM phase fraction at low temperatures and low fields is approximately 45%. As expected, the lower Nd doped sample, y = 0.3, does not show any metamagnetic transition. The M vs. T data of Fig. 2 indicates that this sample is phase separated at intermediate temperatures, and becomes a homogenous ferromagnet below 50 K.

In the remaining of the paper we address several aspects of the metamagnetic transition in $La_{0.125}Nd_{0.5}Ca_{3/8}MnO_3$ (y = 0.5). Figure 5 shows magnetization and specific heat measurements as a function of applied field, at 2.5 and 6.0 K. As observed in the magnetization data of Fig. 5a, the metamagnetic transition is step-like at 2.5 K, and at 6.0 K evolves continuously as the field increases. A similar behavior is observed through specific heat measurements, shown in

Fig 5b. This thermodynamic data yield important insights into the origin of these magnetization jumps. At 2.5 K a discontinuous jump in the specific heat is evident, a clear signature of a first order phase transition. At 6.0 K, the evolution of the specific heat as a function of applied field is smooth. At both temperatures, a large hysteresis between the field increasing and field decreasing branches is observed. It is worth noting that at 2.5 K the specific heat of the upward branch starts above the downward branch, and the opposite occurs at 6.0 K. Therefore, there is an excess specific heat, and therefore an excess entropy, when cooling to 2.5 K, which is released in the magnetization jump. This excess entropy is caused by quench disorder within the low temperature metastable state, associated with the block states previously discussed.

An alternative way to probe the metamagnetic transition is through direct measurements of the sample temperature while the magnetic field is increased, T(H). The results are shown in Fig. 6. With a starting temperature of 2.5 K (Fig. 6a), a very sudden temperature overshoot is observed at a critical field, coinciding with field of the avalanche transition. The sample temperature increases almost instantaneously by approximately 20 K, and this excess heat is subsequently released to the environment. This indicates that the avalanche is triggered by a local and sudden increase of temperature, which transforms non-FM regions in to the FM state. The local increase of temperature helps the neighboring frozen states to overcome the energies barriers, and, in turn, to become unblocked, igniting the avalanche-like process. In Fig 6b we show the T(H) results for a starting temperature of 6 K. The field induces only a small increase in temperature, of less than 2 K, which develops in a wide field range within the metamagnetic transition.

These results establish that the absorbed and released heat play a major role in the metamagnetic transition. There are two competing processes: heat input due to an intrinsic microscopic process and heat released through a thermal link to the outside thermal bath. With this in mind, we performed an experiment which allow us to probe the role of the thermal coupling between the sample and the surrounding heat reservoir in the metamagnetic transition. The results are displayed in Fig. 7. In the standard measuring mode the sample is placed in a low pressure exchange gas, and a magnetization jump is observed. Alternatively, when the sample is placed in liquid Helium, the thermal conductance to the surrounding is much enhanced, and the metamagnetic transition becomes smooth and continuous. The released heat in the microscopic phase transformation easily flows to the outside, with not enough time to generate a heat burst within the sample. These results confirm that the thermal conductance within the sample and how fast heat can be released to the outside environment are crucial parameters which determine the observation of a broad or sharp metamagnetic transition [28]. This is consistent with the field sweep rate dependence previously reported [18,21]. A quantitative analysis would need to take in to account the dimensions of the sample (surface to volume ratio), the internal conductance within the sample, the conductance between the sample surface and the outside environment, and the characteristic times involved, of the order of milliseconds [29]. This is beyond the purpose of this investigation.

**4 – Conclusions**

In this paper we presented an experimental investigation of the avalanche-like metamagnetic transition in polycrystalline $La_{5/8-y}Nd_yCa_{3/8}MnO_3$ manganite compounds. In samples with higher Nd content (y = 0.4 and 0.5), an avalanche-like magnetization jump is observed in M vs. H measurements at low temperatures (2.5 K). The whole sample changes abruptly from a low field state, with a majority of CO-AFM phase, to a high field homogeneous FM state. The magnetization data, at fields below the metamagnetic transition, is consistent with the presence of a metastable blocked state, arising from frozen disorder within the phase separated state. At a base temperature of 2.5 K the metamagnetic transition is discontinuous in both M(H) and C(H) data. Direct measurements of the sample's temperature while the field is increased show a large temperature rise (≈ 20 K) concurrent with the magnetization and specific heat jump. This is consistent with the scenario in which the resulting heat arising from the phase transformation triggers a chain reaction which causes the avalanche. At a slightly higher temperature, T = 6 K, the metamagnetic transition becomes continuous as a function of applied field. The field induced temperature rise is small, and, like the specific heat, continuous along the transition. This behavior is understood by considering the main role of the latent heat. The temperature dependence of the specific heat follows approximately a $T^3$ dependence, and a small base temperature increase gives rise to a huge increase in the specific heat. With the same local heat generated, higher C values yield a small temperature variation, insufficient to ignite the avalanche process. This is why discontinuous jumps are only observed at very low temperatures, and the effect is replaced by a smooth evolution of the magnetization or specific heat a few degrees above. Finally, we showed that the abrupt or continuous character of the transition is determined not only by temperature but also by the thermal coupling of the sample to the surrounding environment. There must be enough time for the heat wave to ignite an abrupt process. A strong thermal coupling to the outside bath prevents the avalanche, and yield a continuous metamagnetic transition.

In conclusion, the close interplay between dynamical and quenched disorder, intrinsic to the phase separated state, coupled to thermodynamic properties relevant to heat propagation, are key factors in the fine tuning governing the phenomenon of avalanche-like metamagnetic transitions in manganites. It is a process triggered by self-heating, which, under specific circumstances, give rise to a very sharp (in field and time) macroscopic spin reversal, due to a phase transformation from CO-AFM to a FM state. Yet, more studies are needed to reveal the precise microscopic mechanism behind this exotic discontinuous transition.

**Acknowledgments**

This research was supported by the bilateral cooperation project FAPERJ-CONICET, by the Brazilian agencies CNPq and CAPES, and by the Brazilian Synchrotron Light Laboratory (LNLS), using the XPD beam line.

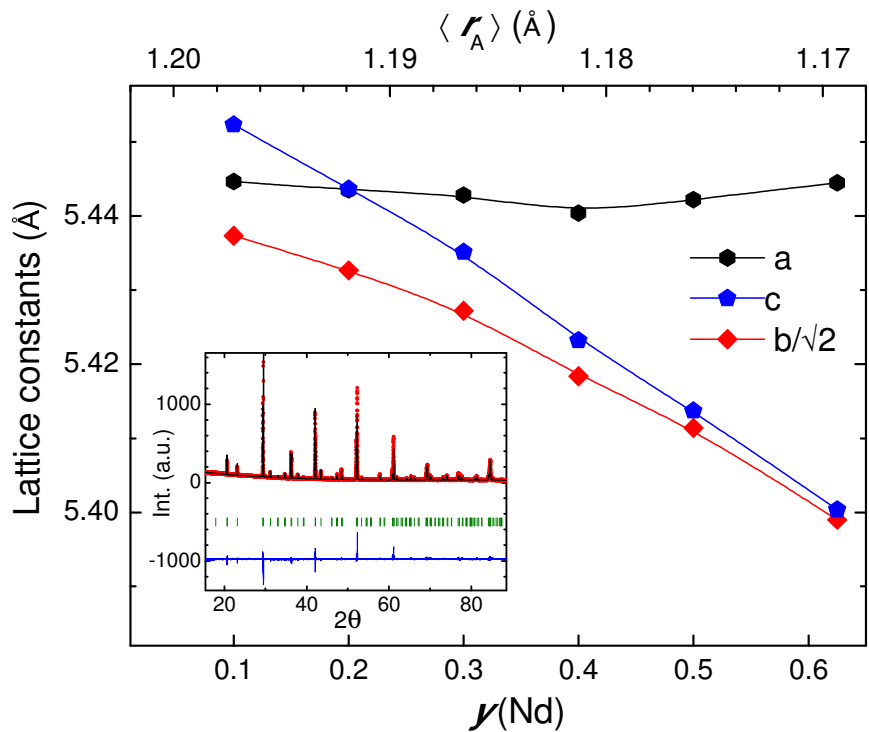

Fig. 1 - Room temperature lattice parameters of $La_{5/8-y}Nd_yCa_{3/8}MnO_3$, plotted as a function of Nd doping $y$, as determined from x-ray diffraction. The top axis shows the average ionic radius of the A-site. The inset shows the diffraction data, Intensity vs. 2θ, for the y = 0.5 sample. The observed and calculated patterns are both plotted. The difference curve is shown at the bottom. Vertical bars indicate the expected Bragg peak positions

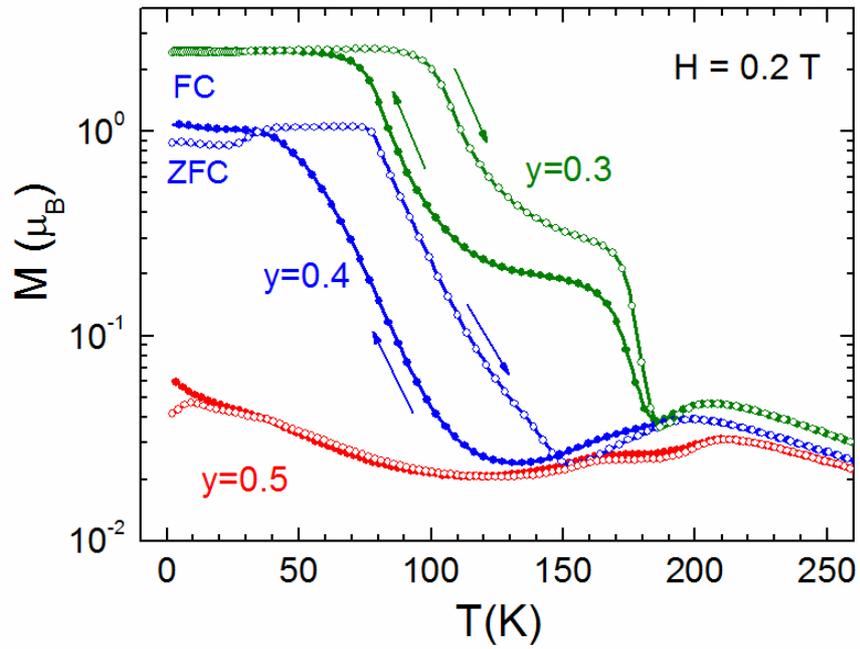

Fig.2 - Zero field cooled (ZFC) and field cooled (FC) magnetization as a function of temperature, for samples of $La_{5/8-y}Nd_yCa_{3/8}MnO_3$, with y = 0.3, 0.4, and 0.5, measured with H = 0.2 T.

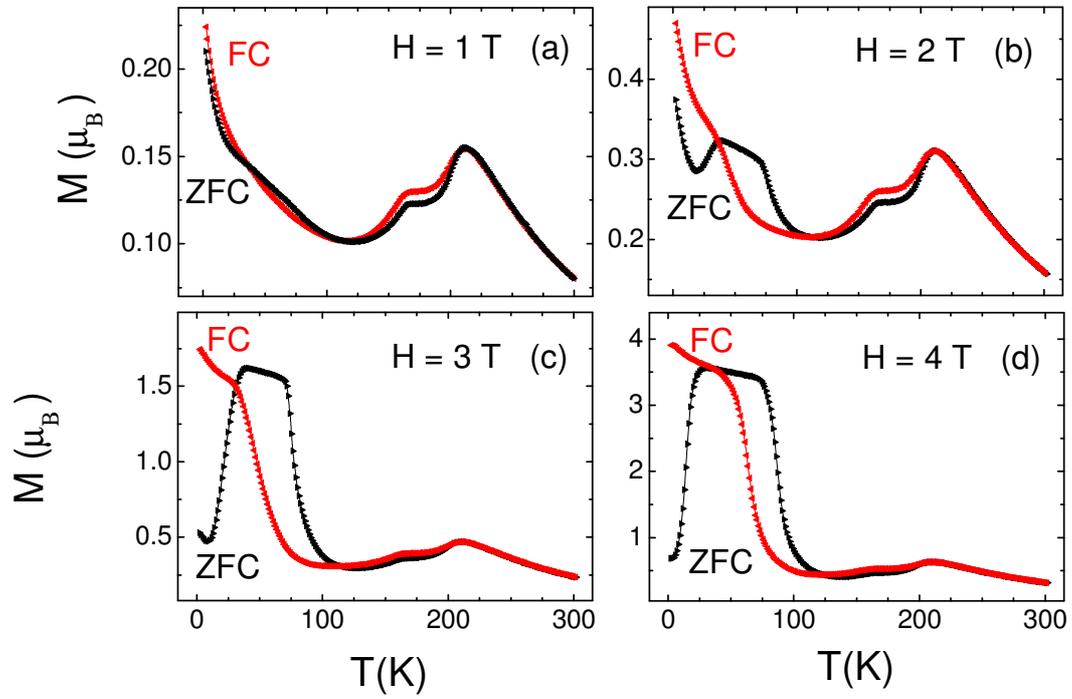

Fig. 3 - Temperature dependence of the zero field cooled (ZFC) and field cooled (FC) magnetization, of $La_{0.125}Nd_{0.50}Ca_{0.375/8}MnO_3$ (Nd 0.5), measured with H = 1, 2, 3, and 4 T.

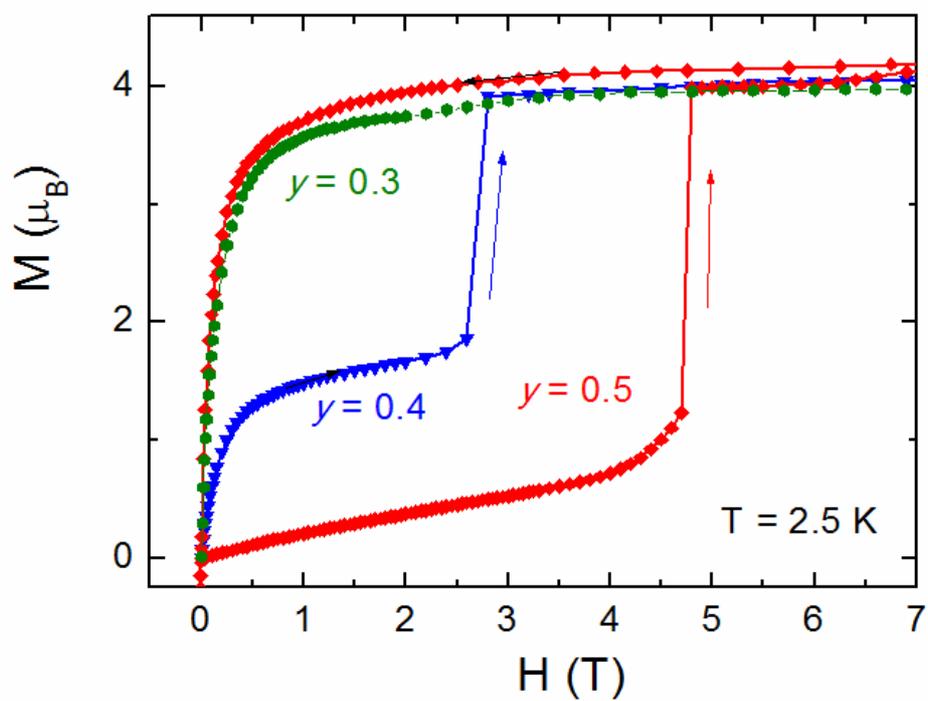

Fig.4 - Magnetization as a function of applied field, for samples of $La_{5/8-y}Nd_yCa_{3/8}MnO_3$, with y = 0.3, 0.4, and 0.5, measured at 2.5 K

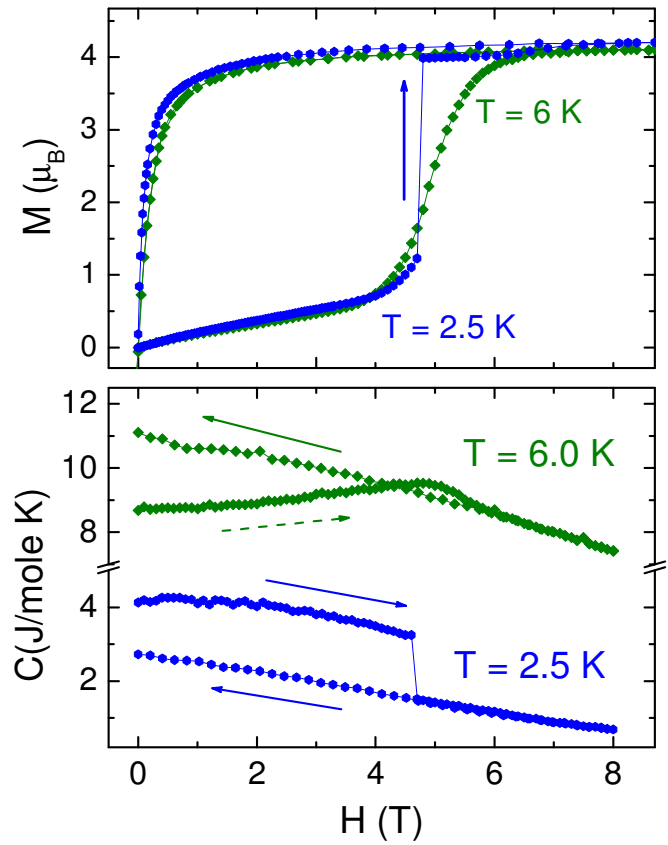

Fig. 5 - Field dependence of the (a) magnetization and (b) specific heat of $La_{0.125}Nd_{0.50}Ca_{0.375/8}MnO_3$ (Nd 0.5), measured at T = 2.5 and 6.0 K.

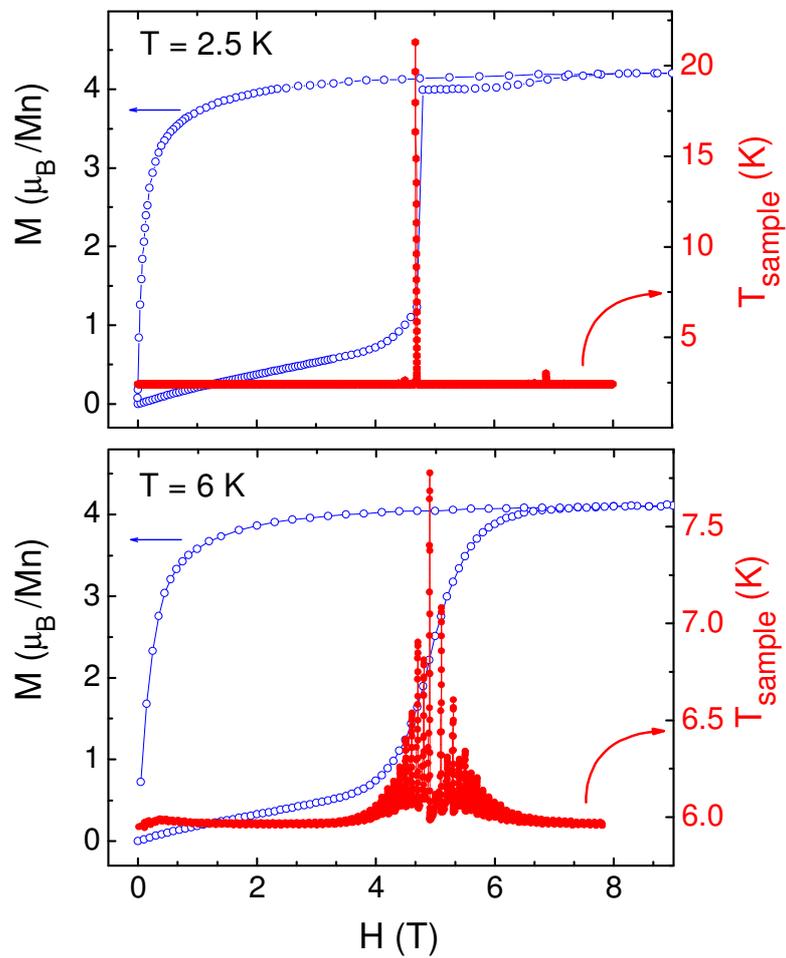

Fig. 6 - Field dependence of the magnetization of $La_{0.125}Nd_{0.50}Ca_{0.375/8}MnO_3$ (Nd 0.5), and direct measurement of the sample temperature, measured at a base temperature T = 2.5 and 6.0 K.

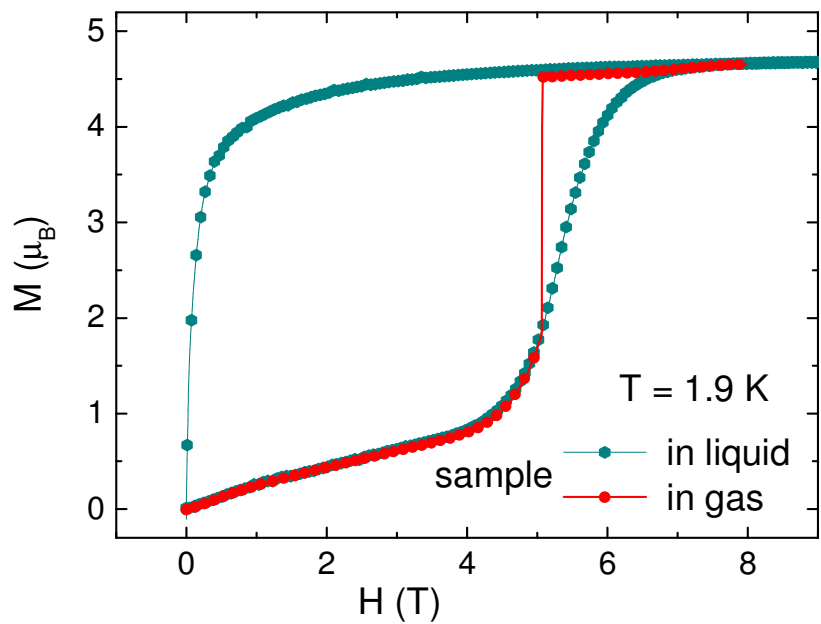

Fig. 7 - Field dependence of the magnetization of La$_{0.125}$Nd$_{0.50}$Ca$_{0.375/8}$MnO$_3$ (Nd 0.5), measured at T = 1.9 K with the sample placed in liquid Helium or gas.